\documentclass[sigconf]{acmart}
\copyrightyear{2022}
\acmYear{2022}
\setcopyright{rightsretained}
\acmConference[IUI '22]{27th International Conference on Intelligent User Interfaces}{March 22--25, 2022}{Helsinki, Finland}
\acmBooktitle{27th International Conference on Intelligent User Interfaces (IUI '22), March 22--25, 2022, Helsinki, Finland}\acmDOI{10.1145/3490099.3511147}
\acmISBN{978-1-4503-9144-3/22/03}

\usepackage{subcaption}
 \usepackage{xcolor}

\begin{document}

\title{NewsPod: Automatic and Interactive News Podcasts}

\author{Philippe Laban}
\email{phillab@berkeley.edu}
\orcid{xx}
\affiliation{%
\institution{UC Berkeley}
\country{United States of America}
}

\author{Elicia Ye}
\email{eliciaye@berkeley.edu}
\affiliation{%
\institution{UC Berkeley}
\country{United States of America}
}

\author{Srujay Korlakunta}
\email{srujay@berkeley.edu}
\affiliation{%
\institution{UC Berkeley}
\country{United States of America}
}

\author{John Canny}
\email{canny@berkeley.edu}
\affiliation{%
\institution{UC Berkeley}
\country{United States of America}
}

\author{Marti A. Hearst}
\email{hearst@berkeley.edu}
\orcid{0000-0002-4346-1603}
\affiliation{%
\institution{UC Berkeley}
\country{United States of America}
}

\begin{abstract}
    News podcasts are a popular medium to stay informed and dive deep into news topics. Today, most podcasts are handcrafted by professionals. In this work, we advance the state-of-the-art in automatically generated podcasts, making use of recent advances in natural language processing and text-to-speech technology. We present NewsPod, an automatically generated, interactive news podcast. The podcast is divided into segments, each centered on a news event, with each segment structured as a Question and Answer  conversation, whose goal is to engage the listener. A key aspect of the design is the use of distinct voices for each role (questioner, responder), to better simulate a conversation. Another novel aspect of NewsPod allows listeners to interact with the podcast by asking their own questions and receiving automatically generated answers. We validate the soundness of this system design through two usability studies, focused on evaluating the narrative style and interactions with the podcast, respectively. We find that NewsPod is preferred over a baseline by participants, with 80\% claiming they would use the system in the future.
\end{abstract}

\keywords{podcast, interactive podcast, automatic podcast, news podcast, conversational, summarization, question answering, question generation}

\maketitle

\section{Introduction}

\begin{figure}
    \centering
    \includegraphics[width=\linewidth]{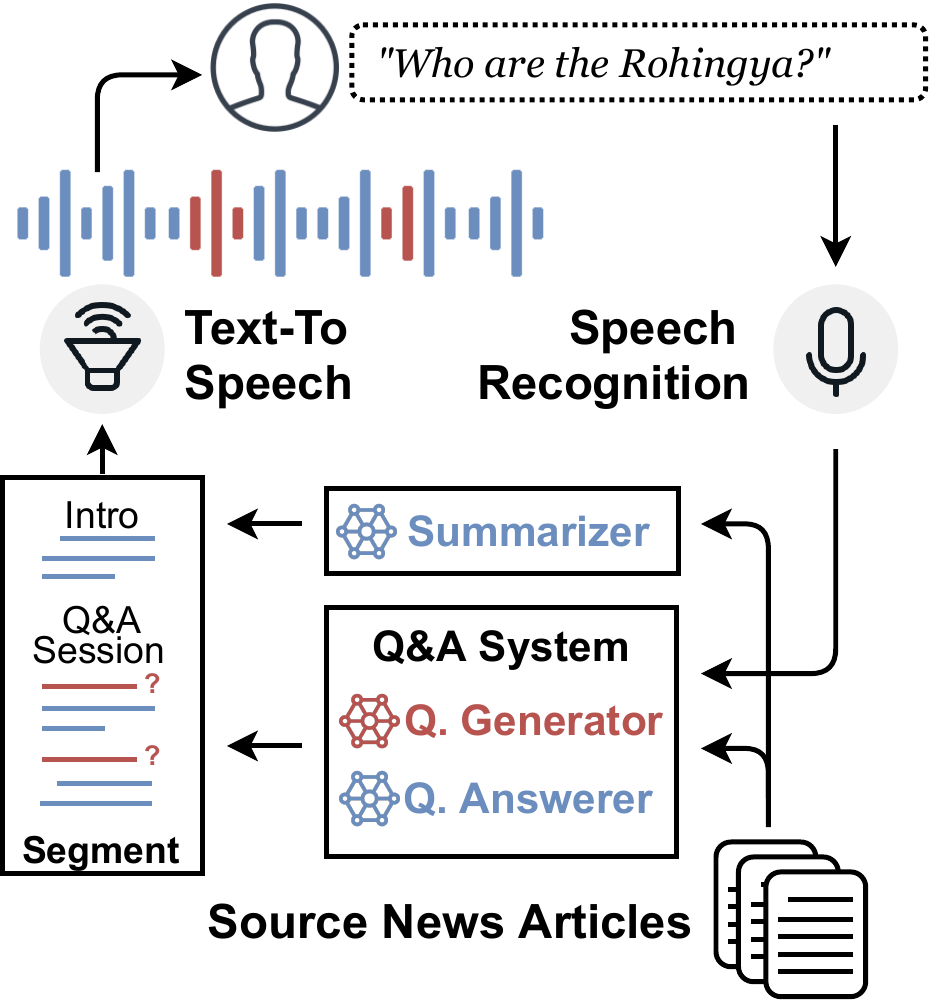}
    \caption{\textbf{In NewsPod, a story is divided into a introductory summary, and an automated Question-and-Answer (Q\&A) session.} The listener can interact with the podcast by asking custom questions. The system integrates an answer into the podcast when one is available.}
    \label{fig:overall_diagram}
\end{figure}

In a given week, only 4 in 10 Americans reported that they delved deeper into a particular news story beyond the headlines, according to a report from the American Press Institute \cite{rosenstiel_media_insight_project_2014}. This statistic is partly explained by the fact that social media has become a significant part of news consumption habits, but also because reading long texts is challenging for many, with 21\% of Americans aged 16 years or older having low English literacy skills \cite{desjardins2013oecd}.

However, 57\% of U.S. adults obtain some of their news through an audio-only platform, a combination of radio and podcast consumption \cite{barthel_radio_2020}. Today, more than 53 million U.S. adults own a Smart Speaker, with 62\% using the speaker one hour a week or more to listen to the news ~\cite{smart_audio_report_2020}.
Given this, we hypothesize that novel, more engaging audio-based interfaces, such as interactive podcasts, hold promise to bring in-depth news engagement to a wider audience.

News podcasts and radio shows are most commonly handcrafted, and curated for the audio platform, often requiring a team of professionals. For instance, a typical episode of \textit{The Daily}, a popular podcast from The New York Times, involves a team with a host, three producers, two editors and an engineer \cite{nytimesdaily}.

On the other end of the spectrum, automation in speech synthesis has progressed rapidly in the last decade.
Recent advancements in text-to-speech with WaveNet \cite{oord2016wavenet} and Tacotron 2 \cite{shen2018natural} have created an opportunity for a new generation of automated speech interfaces.
Automated speech technology has become widely available, for instance through the Voice Browser Working Group of the W3C\footnote{https://www.w3.org/Voice/}, equipping most modern browsers with standardized text-to-speech and speech recognition interfaces.
Natural language understanding (NLU) and generation (NLG) have also seen accelerated development, with the state-of-the-art sometimes found to match human-performance on benchmarks for tasks such as summarization \cite{Zhang2020PEGASUSPW} and question answering \cite{zhang2021retrospective}.

Some news publications, such as \textit{The Economist} and \textit{Bloomberg News}, have leveraged this new technology to create audio versions of their printed news stories and added \textit{playback} options, to enable reading on the go. To generate a playback, a news article is processed unchanged through a text-to-speech engine to produce an audio article played for its audience.
A playback is inherently limited, because the content is not adapted for the audio format, when compared to a manually recorded podcast.

In this paper, we present the first automatic attempt at bridging the gap between automated news reading and curated podcasts.
We present NewsPod, a system to automatically generate news podcasts. NewsPod uses state-of-the-art text summarization and question generation to create a script consisting of an introductory summary, followed by a Question and Answer (Q\&A) session involving multiple automated voices. The listener can interact with the podcast by selecting stories of interest, listening to segments in a non-linear order, and asking open-ended questions during the podcast, which the system attempts to answer automatically.

We demonstrate the effectiveness of NewsPod through two usability studies. The first study focuses on evaluating the novel narrative style, which is compared to two automated baselines, and a manually created reference. The second study measures how participants interact with the podcast, and whether including periodic breaks can encourage participant interaction.

Our work makes the following contributions:
\begin{itemize}
	\item A system called \textbf{NewsPod} that automates on-demand news podcasts, featuring multiple voices to simulate a conversation, and allowing the listener to join the conversation by asking questions,
	\item An evaluation of the narrative format proposed in NewsPod, finding that the conversational podcast we proposed is preferred over a non-conversational baseline,
	\item An evaluation focused on the interactions between the listener and the podcast, finding that when given the opportunity, many participants interact by asking questions; additionally, an analysis of the questions asked reveals limitations in current automatic question-answering technology.
\end{itemize}

A live demonstration of the NewsPod system is available at \url{https://newspod.github.io/}.

\section{Related Work}
\label{sec:related_work}

\textbf{Automating News Podcasts}. To the best of our knowledge, there is no prior work automating the creation of news podcasts. The system of \citet{yoshino2015conversational} is the closest conceptually. They present a chat dialogue system using news sources as the subject matter. The system proposes topics to the listener, who can ask questions about the topic. A key conceptual difference with our work is that \citet{yoshino2015conversational}'s system is designed for open-ended information navigation, requiring user input to advance the conversation, while we prepare a script for a podcast targeting a specific duration, allowing for a combination of passive listening or active question asking, with listener interruptions allowed at any point in the conversation. Another important difference is the use in our work of multiple synthesized voices to simulate a conversation.

Some authors have written position papers and made forecasts about a future that includes automatically generated and personalized podcasts as is done in NewsPod.
For instance, \citet{Dubiel2019InquisitiveMA} suggest that news presented in a conversational audio form could increase engagement, allow for the presentation of multiple viewpoints, and allow listeners to ask follow-up questions to expand their knowledge on the subject. The BBC's R\&D laboratory has published a series of vision papers on the future of audio broadcasting, including forecasts for a future in which audio news topics and lengths can be personalized \cite{bbc_audio_2019}.

\citet{Lochrie2018DesigningIA} manually built prototypes of news podcasts and performed an evaluation, which they used to propose seven recommendations for automating news podcasts. NewsPod follows four of these recommendations, namely: (1) the content should be adapted to the audio format with bite-size chunks of information (in NewsPod: questions and answers), (2) summarize key points, (3) leverage several specialized voices, (4) adapt the narrative style to the context. 

\textbf{Conversational Interfaces for News Recommendations.} 

\noindent
Several systems have been proposed that retrieve news articles in response to a user's queries within the format of an interactive dialogue. \citet{bechberger2016personalized}  
retrieve a news event based on a user query and the user's general interests, using text chat rather than audio.
\citet{Sahijwani2020WouldYL} investigate different approaches for recommending news content as part of an Alexa Prize conversational agent. Their system suggests a news article as part of the conversation, and if the user accepts the recommendation, a search engine is queried for news on an entity or a trending topic until the user rejects the recommendation.
They find that reducing the formality of the recommendation and increasing the specificity leads to significantly higher acceptance rate from listeners.

\textbf{The multiple source effect} \cite{Harkins1981TheMS} posits that arguments coming from multiple distinguishable sources increases persuasion when compared to similar arguments made by a single source. Experiments have also validated the multiple source effect for video \cite{Harkins1981EffectsOS} as well as audio with synthetic speech \cite{lee2004multiple}. Most related to our work, \citet{Kang2012EffectsOP} find that when visitors of a national park listen to a podcast tour, including multiple voices in the podcast positively impacts the social presence and enjoyment of the visitor. In NewsPod, we capitalize on the multiple source effect by dividing the content amongst several roles, each synthesized by a distinct automated voice.

\textbf{Conversational Narration Style}. Prior work has shown that modifying the narration style of multimedia content can have an impact on the listener's understanding, retention, and enjoyment. Personalization by altering the point of view from neutral (3rd-person) to conversational (1st- and 2nd-person) has been shown to be effective in many learning domains in reducing cognitive load and improving listener understanding \cite{Moreno2000EngagingSI, Mayer2004APE, Ginns2010PersonalizationEL}. In their study, \citet{Kang2012EffectsOP} also find that when podcast tours follow a conversational format, it has a positive impact on the listener's enjoyment and social presence. By simulating a conversation in NewsPod with voices asking questions answered by others, we attempt to create an informal environment that is inviting to the listener.

\section{Background: Natural Language Processing}

We provide relevant background on recent progress in the field of natural language processing (NLP). We leverage several of these components in creating NewsPod;  their capabilities and limitations shape our choices in the design, as well as the interaction possibilities for users of the system.

\subsection{Question Answering}

Recently, large neural-network based models \cite{devlin2019bert,clark2019electra} have shown remarkable performance at question-answering, most commonly in the \textit{extractive question answering} setting: given a paragraph and a question, extract a span in the paragraph (e.g., an entity, a phrase, etc.) that answers the question. Additionally, models must provide their confidence in the answer, or else predict that the paragraph does not contain an answer to the question \cite{rajpurkar2018know}.

With models outperforming human performance on answer extraction on the standard (albeit limited) dataset benchmarks \cite{zhang2021retrospective}, the community has proposed more challenging settings for question answering. We briefly describe two relevant to our work: \textit{conversational question answering} and \textit{open-domain question answering}.

In conversational question answering projects such as CoQA \cite{Reddy2019CoQAAC} and QuAC \cite{choi2018quac}, a model must answer a user's question as part of a longer, contextualized sequence of questions. This is more challenging than one-off Q\&A, because the model must be able to handle phenomena such as context co-reference, and follow-up questions, removing the strong assumption that a question can be answered correctly regardless without prior context.

In open-domain question answering, a model is given a question and must first retrieve relevant paragraphs from a large corpus of text before attempting to answer the question \cite{Moldovan2000TheSA}. Commonly the corpus of text is very large (e.g., all of English Wikipedia \cite{Chen2017ReadingWT}), creating an engineering challenge in retrieving the needle of relevant paragraphs from the large haystack of paragraphs in the text corpus. Even though recent work \cite{Karpukhin2020DensePR} has made great strides in making open-domain question answering computationally tractable, it still requires specialized equipment (i.e., several GPUs) and data-storage software.

In NewsPod, we use a standard extractive Q\&A model, a RoBERTa model trained on common extractive Q\&A datasets, which can base its answers on a limited set of relevant news articles. This choice's main advantage is its low computational requirement and simplicity in setting up, at the cost of limiting the types of questions the system can answer to questions answered in news articles. We discuss this choice further in Section~\ref{sec:study_interaction}.

\subsection{Summarization}

Abstractive summarization has become one of the most active research areas in automatic text generation, and is a common application of large language models. Typically the models take as input a single text document, and produce a summary by generating it one word at a time from a fixed vocabulary. Most models require as training data large datasets of aligned (article, summary) pairs, and so the NLP community has curated a diverse family of datasets as benchmarks, ranging from legal texts \cite{Kornilova2019BillSumAC,Sharma2019BIGPATENTAL} to research papers \cite{cohan2018discourse} to online forums (i.e., Reddit) \cite{kim2018abstractive}.

In the news domain, the two most common datasets are CNN\-/Daily Mail \cite{Hermann2015TeachingMT} and Newsroom \cite{grusky2018newsroom}. The latest models, such as PEGASUS \cite{Zhang2020PEGASUSPW}, have shown ability to generate summaries that achieve high quality on attributes of fluency and informativeness, validated through human assessments.

However, a known major limitation of automated abstractive summarization \cite{Huang2020WhatHW} is the absence of a guarantee that generated summaries are factual: the model can ``hallucinate'' content or otherwise not faithfully represent the contents of the document it summarizes. Although the community has proposed methods to optimize factual correctness \cite{Zhang2020OptimizingTF}, the absence of a guarantee limits the usability of neural summarization at large.

In NewsPod, we use the PEGASUS model, as it is publicly available\footnote{https://huggingface.co/models?search=pegasus}, and is currently the state-of-the-art for news summarization. In some cases in our dataset, a news source provides a journalist-written summary alongside with the full body content. When a human-written summary is available, we prioritize its use, and produce model-generated summaries in the absence of a human-written one.

 \subsection{Question Generation}
 
In question generation \cite{du2017learning,Pan2019RecentAI}, a text generation model is usually given a paragraph and tasked with generating a question that can be answered by the paragraph. Question generation is generally under-constrained, as a paragraph typically answers many valid questions.
In under-constrained situations, text-generation models favor most frequent outputs, which can be generic and uninteresting in the case of question generation (e.g., \textit{What happened?}).
To counter this, a standard strategy is to constrain question generation to target a specific answer for the question or constrain the model to use a desired first word (i.e., How, Why, etc.), encouraging the model to generate non-generic questions.

Beyond specificity of questions, recent analysis of expert annotations \cite{murakhovs2021mixqg} of automatically generated questions reveal three categories of error for QGen models. First, questions can lack fluency and naturalness (e.g., \textit{What is the person ...} instead of \textit{Who is the person ...}). Second, the model can generate questions that are off target, meaning that the questions are not answered by the input paragraph intended to contain answer. Third, questions might be poorly phrased for the given context, focusing on unimportant aspects of the paragraph (e.g., a specific figure or the name of a person mentioned).

In NewsPod, we use a standard GPT2 model finetuned on a standard QGen dataset. In order to account for these limitations, we generate a large number of questions, generating 7 questions for each paragraph. We obtain many more questions than we require, and use a graph-based method to select questions that are most relevant.

\section{Demonstration of NewsPod}

To convey the experience of using NewsPod, we describe a short example scenario. Consider Lee, who has been busy with a deadline recently and is interested in getting caught up on the news.

\subsection{Selecting Podcast Segments}

Lee opens the NewsPod interface, and first selects a target duration of 5 minutes for the podcast, to get a quick update on latest topics of interest.

Once the target duration is selected, Lee personalizes the podcast by choosing the segments to include, selecting three segments about: (1) a Tesla self-driving ban, (2) the Rohingya crisis, and (3) an Amazon Union Vote. Upon Lee's selection, the podcast starts with a greeting, and transitions into the first segment.

\subsection{Segment Reading}

The Tesla Self Driving ban segment begins with the reading of a recent headline, and a summary of the latest developments (selected from a CNBC news article):
\begin{quote}
    \textcolor{blue}{ \textbf{[Voice 1]}}``Tesla uses customers to test unfinished driverless tech, warns NTSB, urging stricter standards.
    In a letter last month, NTSB chief Robert Sumwalt named Tesla 16 times while urging the NHTSA to set stricter standards.''
\end{quote}

Lee is unfamiliar with some of the acronyms, and decides to open the Transcript panel to read through the agency names while the podcast begins the Q\&A Session.

During the Q\&A Session, Lee gets distracted by an email notification and loses track of what is being said. A change of speaker in the podcast segment asking a question refocuses Lee's attention: 
\begin{quote}
    \textcolor{red}{ \textbf{[Voice 2]}}``What would be called a fully autonomous vehicle?''
\end{quote}

The segment answers the question:
\begin{quote}
    \textcolor{blue}{ \textbf{[Voice 1]}} ``Tesla's current system is Level 2 autonomous. Tesla cars have some automated functions, but require drivers to remain attentive and keep their hands on the wheel. The highest level, level 5, would be a completely autonomous vehicle that never requires driver attention.''
\end{quote}
Finally, the segment ends with a quote from Elon Musk:
\begin{quote}
    \textcolor{green}{ \textbf{[Voice 3]}}``I think Autopilot's getting good enough that you won't need to drive most of the time unless you really want to.''
\end{quote}

The podcast transitions to the second segment Lee selected about the Rohingya crisis.

\subsection{Interacting with the Podcast}

During the introduction of the second segment, Lee feels unfamiliar with the subject and decides to ask a clarification by clicking on the microphone icon. The podcast pauses immediately, and Lee asks: 

\begin{quote}
    \textbf{[Lee]}``Who are the Rohingya?''
\end{quote}

The system remains silent while it processes the question for a few seconds, and replies:
\begin{quote}
    \textcolor{blue}{ \textbf{[Voice 1]}}``The Rohingya are a stateless Muslim minority in Myanmar's Rakhine State. I got the answer from the following paragraph: The Rohingya are a stateless Muslim minority in Myanmar's Rakhine State, thought to number about 1 million people.''
\end{quote}

Having answered Lee's question, the podcast resumes and proceeds with the segment's Q\&A session. The podcast moves on to the third segment about a vote for a union at Amazon. Lee listens through the segment, and towards the end, clicks on the microphone and asks:
\begin{quote}
    \textbf{[Lee]}``Why would the workers oppose forming a union?''
\end{quote}
The system pauses, and after a few seconds, because it is not able to obtain an answer, it replies:
\begin{quote}
    \textcolor{blue}{ \textbf{[Voice 1]}}``Sorry. I couldn't find the answer. If you rephrase I will try again. Otherwise I'll keep walking you through the segment.''
\end{quote}

Lee does not follow-up in the podcast, but makes a note to discuss this with a friend working at Amazon. The podcast resumes, and concludes with a closing sentence, having reached the target duration of five minutes.

\section{Building the Automatic Podcast}
\label{sec:system_description}

We now describe the components used in automating the NewsPod podcast. Figure~\ref{fig:overall_diagram} gives a high-level diagram of the components and their relations. We focus on applying our procedure to the English news domain, noting that the approach could be extended to other languages, since the majority of components (such as the text-to-speech or the summarizer) have equivalents in other languages.

We first describe the elements of the NewsPod web interface shown in Figure~\ref{fig:newspod_interface_overview}. Second, we outline how scripts for podcast segments are automatically generated by identifying groups of news articles that can form the basis of a podcast segment (\S\ref{section:identify_sections}), and generating a segment's Q\&A session (\S\ref{section:gen_qa_session}). In \S\ref{section:from_tts}, we describe how the script is transformed into audio through text-to-speech. Finally in \S\ref{section:live_qa} we detail the live Q\&A system we developed to respond to listener-prompted questions.

\subsection{Podcast Interface}
\label{section:podcast_interface}

\begin{figure*}
    \centering
    \includegraphics[width=0.9\textwidth]{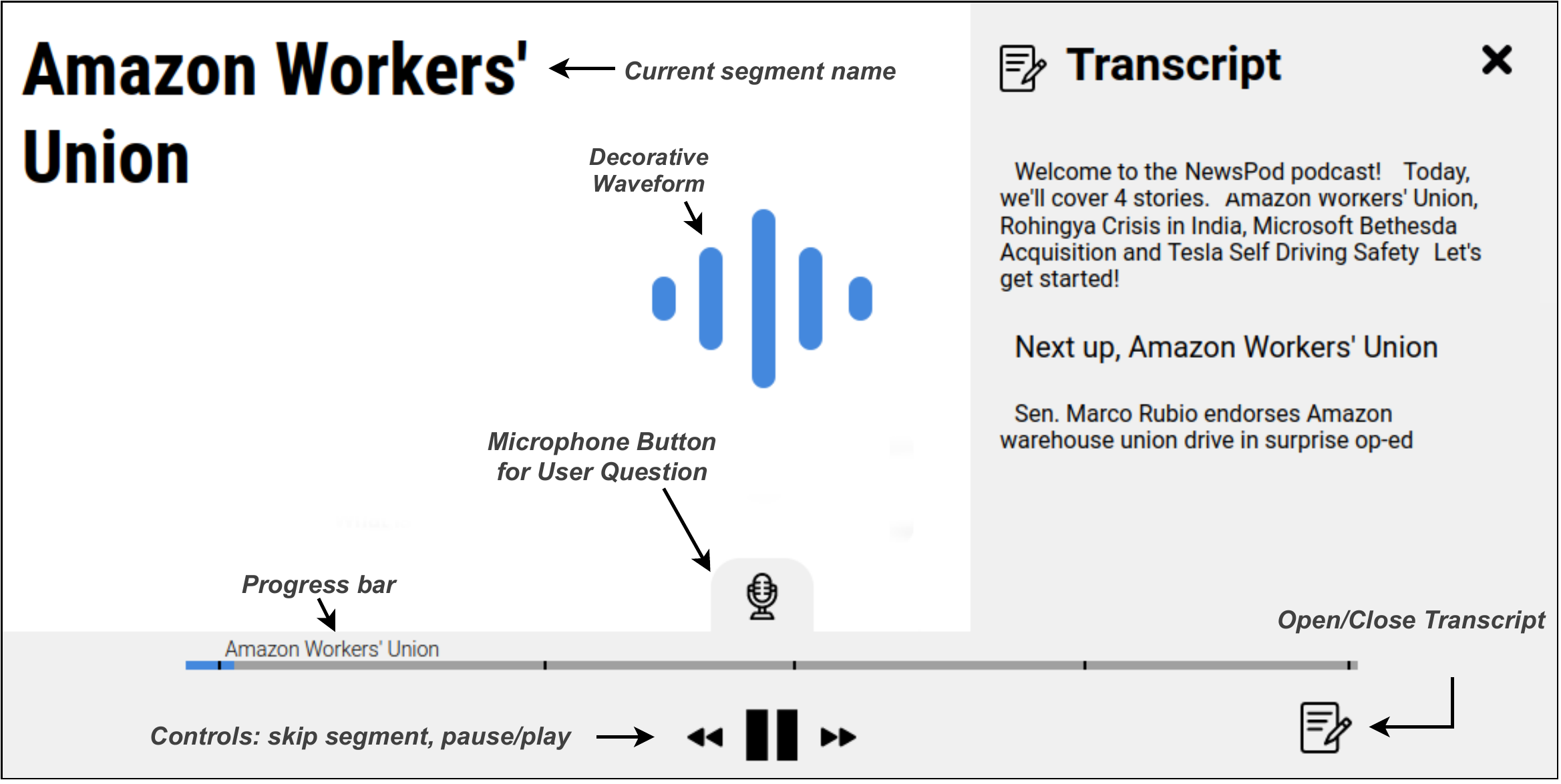}
    \caption{NewsPod's interface resembles other audio playback interfaces: a bottom panel contains common user controls such as pause and play and move forward and backward between podcast segments. The user can toggle a Transcript panel that opens on the right-hand side. The listener can interact with the podcast by clicking on the microphone and asking a question.}
    \label{fig:newspod_interface_overview}
\end{figure*}

A snapshot of the NewsPod graphical user interface (GUI) is shown in Figure~\ref{fig:newspod_interface_overview}. The interface contains three main components: the control panel (bottom), the transcript panel (right), and the decorative wave.

\textbf{The control panel}. The listener has the option to play and pause the podcast, as well as skip forward and backward across segments of the podcast. A transcript button toggles the transcript panel to open and close on the right hand side. Finally, a sectioned progress bar indicates the progression through the podcast, with the beginning of each segment indicated. A user can click on the progress bar, which will redirect the podcast to the start of the corresponding segment. At any time, the listener can click on the microphone button, which will pause the podcast and initiate the live Q\&A procedure outlined in Section~\ref{section:live_qa}.

We considered not requiring the user to click to activate the microphone, and instead continuously monitor the microphone. However, this led to technical difficulty with the microphone sometimes detecting questions from the podcast's voiced text. Continuously monitoring a user's microphone is also concerning from a privacy standpoint \cite{lau2018alexa}, and we opted for a button-based microphone activation.

\textbf{The transcript panel} opens on the right-hand side of the interface, with text from the podcast added one sentence at a time, to allow the listener to follow along in real time as the podcast is playing. The transcript panel is initially closed, and in the usability study of Section~\ref{sec:study_structure}, we analyze what fraction of participants opened the transcript.

\textbf{The decorative wave}, an abstract animated visualization of the generated speech, brings animation to the interface. The wave changes color, turning blue, red and green based on voice identity. When the podcast is paused, the wave flattens and turns grey.

The interface is designed with responsive elements \cite{frain2012responsive} and is compatible with desktop, tablet and mobile environments.

\subsection{Podcast Organization}

In NewsPod, a user can customize the content of the podcast by selecting which topics to include in the podcast and selecting a desired podcast duration. Based on those criteria, a podcast is generated on demand.

The podcast is composed in the following way: (1) a greeting (i.e., ``Welcome to NewsPod, today we'll be covering ...''), (2) a sequence of \textit{podcast segments}, and (3) a closing sentence (i.e. ``That's it for today, thank you for tuning in.'').

Each podcast segment is centered around a particular news topic selected by the user. Importantly, we construct podcast segments in a way that allows for adaptation to user-requested duration. As an example, if a participant selects five segments and a podcast duration of five minutes, each segment must have a duration of one minute. However, if the total podcast duration selected is fifteen minutes, podcast segments can span three minutes each.

\begin{figure}
    \centering
    \includegraphics[width=\linewidth]{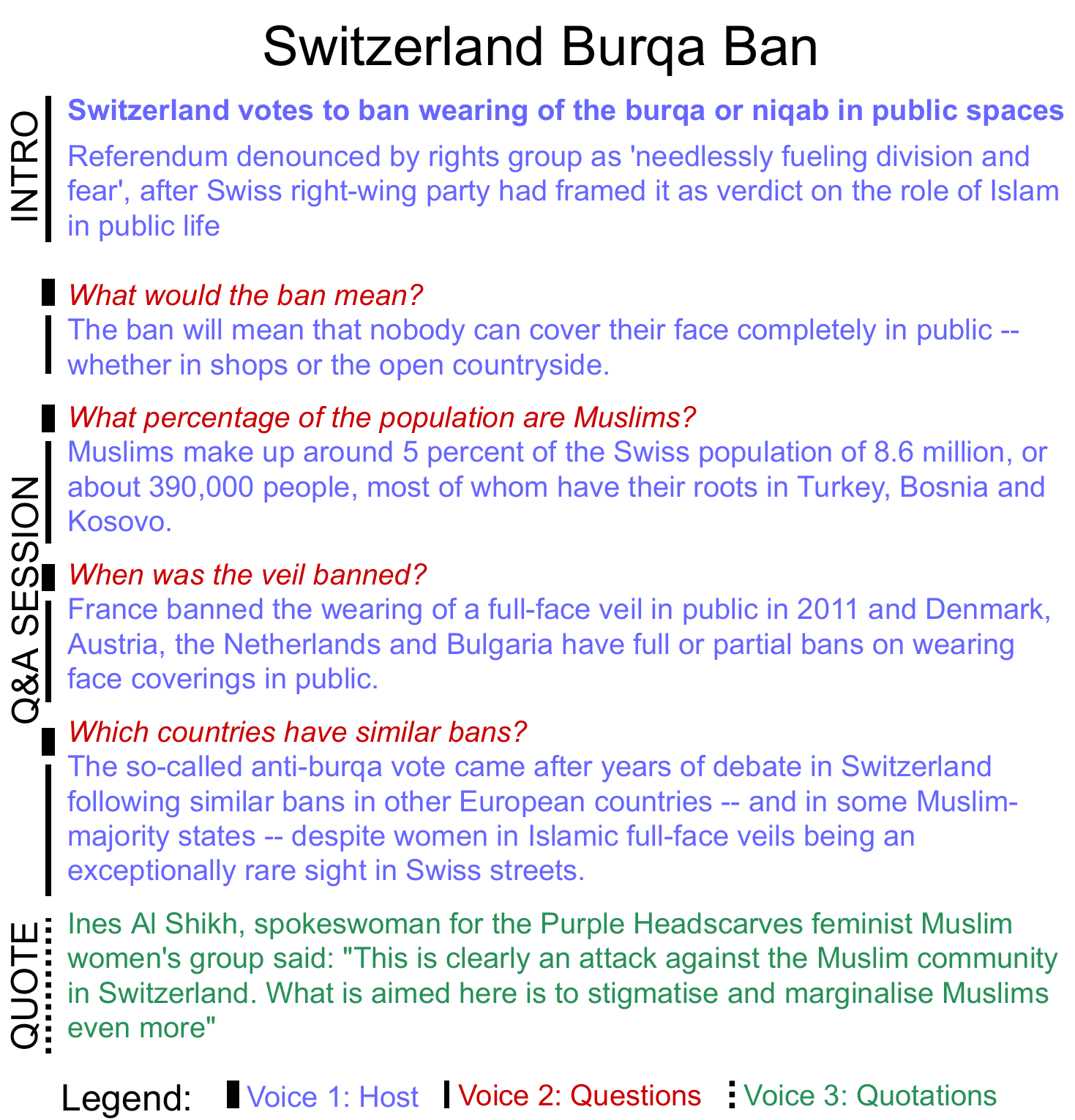}
    \caption{Example generated podcast segment based on six source news articles from \textit{France24, Aljazeera, the Guardian, the BBC, the Middle East Eye and the Times of India}. The segment is composed of an introductory headline and summary, a Q\&A session, and a quote paragraph. Each text color (blue, red, green) indicates a distinct voice.}
    \label{fig:example_segment}
\end{figure}

To accommodate length variability, we construct podcast segments using an \textit{inverted pyramid} writing style common in the news domain \cite{po2003news}. The segment is composed of an introduction consisting of a short headline and a multi-sentence summary, followed by a Q\&A session revealing information in order of most important to most specific. An example podcast segment with a target of 200 words is given in Figure~\ref{fig:example_segment}. It contains a headline, a summary, four question/answer pairs and a quote. If a user desired a podcast half the length, this segment would be truncated after 100 words, revealing only the headline, summary and two question/answer pairs.

\subsection{Identifying Podcast Segments}
\label{section:identify_sections}

In order to create a podcast segment, we first collect a group of related articles discussing a news event, which we call the \textit{source articles}. We then construct an introduction to the segment consisting of a headline and a short summary, with a length requirement of at least two sentences and 20 words. The introduction is meant to give the listener a high-level understanding of the segment before diving into the Q\&A session.

The source article groups are collected using an existing news dataset \cite{laban2017newslens} based on the live feed of around 20 international news sources in English, using an NLP-based clustering algorithm \cite{laban2021news}.

The introductory headline is selected from the available headlines in the source articles based on simple rules that both encourage shorter headlines and discourage the presence of special characters such as ``:'' or ``-'', since these tend to create difficulty in the text-to-speech engine.

The introductory summary can either be selected from an article or generated. A subset of the sources in our dataset provide hand-written summaries; if one of the source articles contains a summary that satisfies the length requirement, we select it. Otherwise we generate candidate summaries using the PEGASUS model \cite{Zhang2020PEGASUSPW}, a neural-network based summarizer, following using a beam search decoding strategy with a beam size of 4. We generate one summary for each source article, and select the summary that both satisfies the length requirement and obtains the highest likelihood according to the model.

\subsection{Generating a Q\&A Session}
\label{section:gen_qa_session}

\begin{figure}
    \centering
    \includegraphics[width=\linewidth]{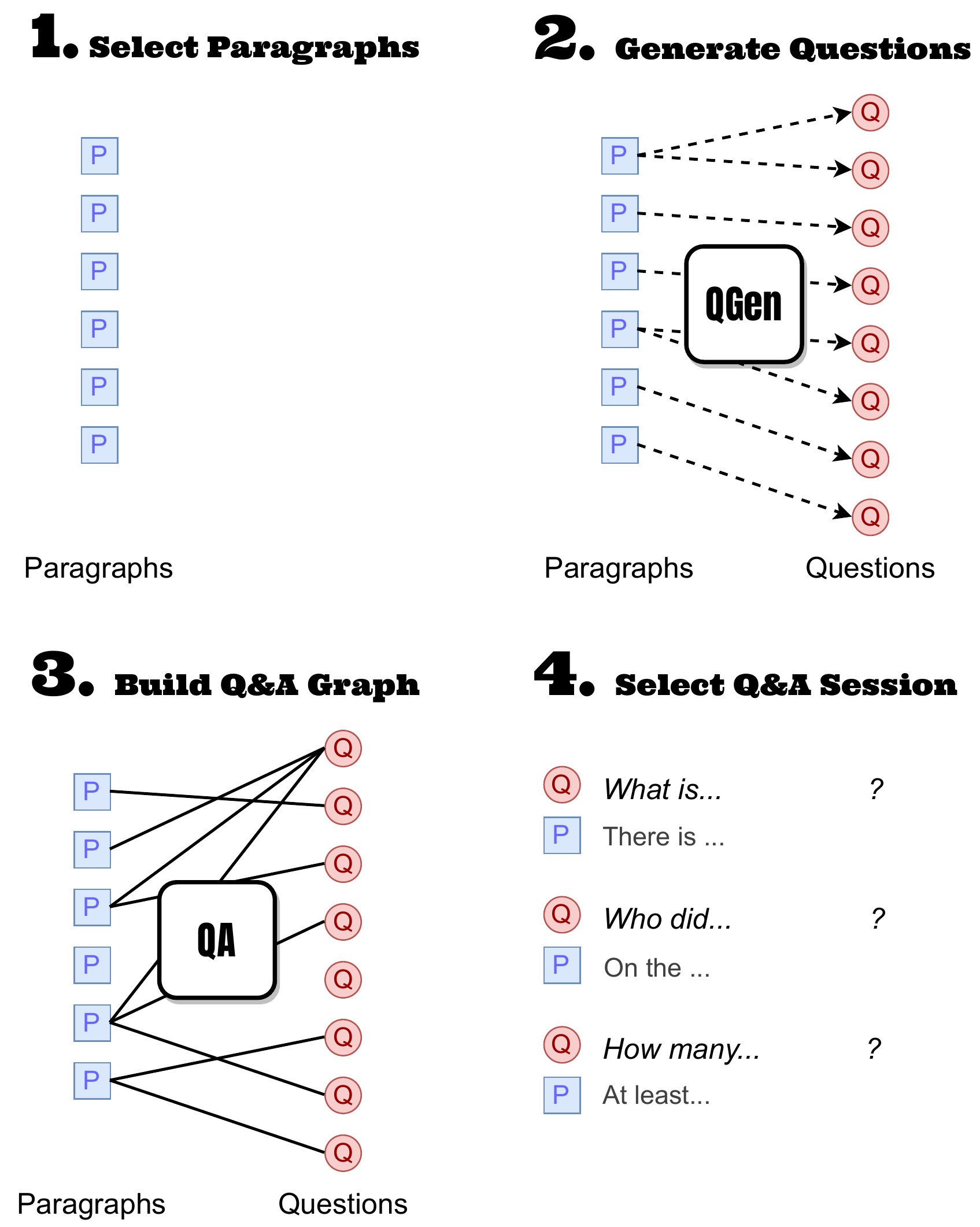}
    \caption{The process of generating a Q\&A session from source articles follows four steps: (1) selecting paragraphs, (2) using a model to generate questions from the paragraphs, (3) using a QA model to detect which paragraphs answer which questions, and (4) finalizing the Q\&A session by ranking and selecting question/paragraph pairs.}
    \label{fig:qa_session_diagram}
\end{figure}

The objective of the Q\&A session is to divide the content into modular pieces involving several voices, with the goal of supporting conversational narrative style and the multiple source effect as described in Section~\ref{sec:related_work}. The questions particularly can serve as a device to focus the listener's attention. We use two neural network models and a graph-based algorithm to generate the Q\&A session out of the news article's content. First a question generator (\textit{QGen}) model produces candidate questions, then an extractive question answering (\textit{QA}) model finds paragraphs containing the answers to the candidate questions, and finally a graph algorithm determines which \texttt{(question, paragraph)} pairs to select for the final Q\&A session.

\textbf{QGen} is a model that takes as input a paragraph of text, and generates a question intended to be answered by the paragraph. We follow current state-of-the-art methodology for QGen models and finetune a large pre-trained language model (a GPT2 model \cite{radford2019language}) on a (paragraph, question) pair dataset (in our case, the NewsQA dataset \cite{trischler2017newsqa}). We finetuned the GPT2 model using standard cross-entropy loss with the ADAM optimizer \cite{kingma2015adam} and a learning rate of $1*10^{-5}$ and a batch size of 24. Training hyperparameters were chosen through tuning on the NewsQA validation portion, with the final model achieving a test perplexity of 4.57.

The \textbf{QA} model takes as input both a paragraph and a question, and produces two outputs: (1) whether or not the answer to the question is in the paragraph, and (2) if it is, which span in the paragraph corresponds to the question. We follow current state-of-the-art methodology and finetune a pre-trained bi-directional Transformer (a RoBERTa-Large \cite{liu2019roberta}) on two common Q\&A datasets: SQuAD V2 \cite{rajpurkar2018know} and NewsQA \cite{trischler2017newsqa}. We use the ADAM optimizer with a learning rate of $2*10^{-5}$ and a batch size of 32. Hyperparameters were selected by tuning on a validation set consisting of the released validation portions of SQuAD V2 and NewsQA. Our final model achieves an F1 score of 86.7 on the SQuAD 2.0 test set, and 68.9 F1 on NewsQA's test set, on par with previously published results \cite{liu2019roberta}.

We now describe how we combine the QGen and QA model to transform the content of several news articles into a Q\&A session. The process, illustrated in Figure~\ref{fig:qa_session_diagram} is divided into four steps: selecting paragraphs, generating questions, building a Q\&A graph, and finalizing the Q\&A Session.

\textbf{Selecting Paragraphs}. The body content of all news articles in the story serves as the basis for the creation of the Q\&A session. In our dataset, a segment is usually constructed based on the content of 6 articles. Articles are automatically divided into individual paragraphs by following end of paragraph character indications (i.e., \texttt{\textbackslash{}n}), yielding 15 paragraphs for each article on average. The paragraphs are further filtered down by removing paragraphs based on two \textit{filtering criteria}: length (keeping only paragraphs with 10 to 45 words), and removing paragraphs with direct quotations, since quotations are processed separately.
As an estimate of the number of paragraphs being processed, a Q\&A session that is being composed from 4 source news articles would be based on approximately 50 paragraphs (if 80\% pass the filtering criteria).

\textbf{Generating Questions}. The QGen model we employ generates questions in an auto-regressive fashion, generating one word at a time in a left to right order. We can ensure that the generated question begins with a specific question word by constraining the model to select a desired word. We use this constraint strategy to generate seven questions for each valid paragraph of content, one starting with each of the following words: \texttt{Who, What, Why, How, When, Where, Which}. When questions across paragraphs, the process often generates on the order of hundreds of questions, which is an order of magnitude above the requirement of a Q\&A session. The next step is to build a graph which can be used to score question relevance to narrow down which questions to include in the Q\&A session.

\textbf{Building the Q\&A Graph}. Once questions have been generated by the QGen model, we use the QA model to find which paragraphs answer each question. The model attempts to answer every question with every paragraph; this is accomplished by building a bipartite graph between the paragraphs and questions. For each \texttt{(paragraph, question)} pair, the QA model assigns scores to a list of potential spans, as well as to a special ``No Answer'' span. If the ``No Answer'' span obtains the highest score, we assume that the paragraph does not answer the question and add no edge to the graph. Otherwise, we select the highest-scoring answer span as an answer to the question and add an edge in the graph between that paragraph and question (see Figure~\ref{fig:qa_session_diagram} for a visual illustration of the graph).

The QA graph has properties of interest: for instance, the degree of the questions indicate how many times a question was answered by distinct paragraphs. The question's degree can therefore indicate whether a question is about a key element of the topic. Vice-versa, the degree of a paragraph roughly indicates how informative it is: how many distinct questions it answers. Prior work \cite{laban2020s} has presented a similar paragraph-question graph construction to track the state of conversation in a chatbot. In NewsPod, we use the properties of this  graph in a different way, to rank question and paragraph relevance.

\begin{figure}
    \centering
\includegraphics[]{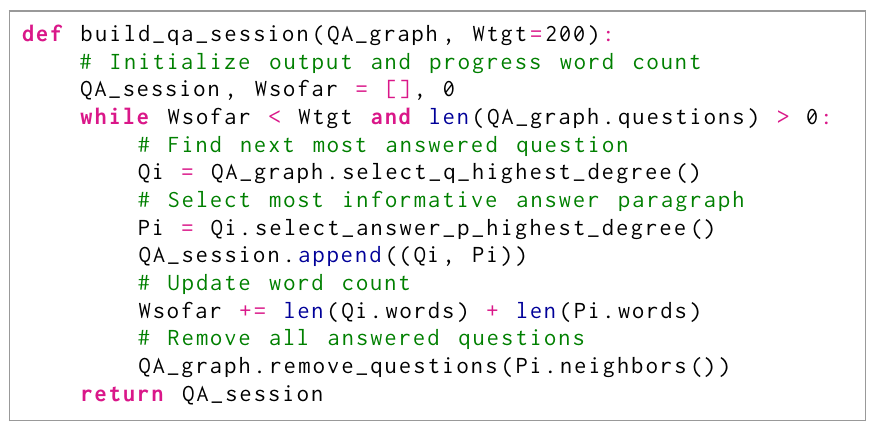}    \caption{Python code for greedy selection of question-answer pairs to generate a Q\&A Session from a Q\&A graph.}
    \label{fig:qa_session_pseudocode}
\end{figure}

\textbf{Selecting the Q\&A Session.}
Figure~\ref{fig:qa_session_pseudocode} provides Python code for the procedure to generate the Q\&A Session from the Q\&A Graph, also taking as an input parameter \texttt{Wtgt}, a target length in number of words. At a high level, we greedily select questions that are answered the most times (an approximation for relevance). Once the top-ranked question is selected, the algorithm selects a paragraph that answers this question, creating a (question, paragraph) pair. It then removes all questions from the graph that this paragraph answers, to limit repetition of content, and iterates until reaching a target word length.

We expect that two undesirable properties --- lack of coherence and repetition  --- may be present in the Q\&A Session. In Section~\ref{sec:study_structure}, we perform a usability study to evaluate the viability of our content selection procedure, comparing both automatic and hand-written podcast sessions.

\textbf{Processing Quotations.}
We use a simple rule-based method to isolate paragraphs that contain direct quotations. We purposefully exclude quotation paragraphs from the Q\&A session on the assumption that quotations more directly relate to a person or organization's opinion than to the answer of a factual question. We use the spaCy dependency parser model \cite{Honnibal_spaCy_2020} to extract three components for each quotation paragraph: the author of the quote, a description of the speaker, and the actual quote.
If at least one quote is successfully extracted, the paragraph containing the quote is added to the end of the podcast segment. If multiple quotes are available, the quote from the most mentioned speaker is selected. Because of its more subjective nature, we position the quotation paragraph as the last element of the podcast session.

An example of a generated segment, including the introduction, the Q\&A session and an extracted quotation is given in Figure~\ref{fig:example_segment}. More segment examples in Appendix~\ref{appendix:example_podcast_segments}.

\subsection{From transcript to speech}
\label{section:from_tts}

\begin{figure}
    \centering
    \includegraphics[width=\linewidth]{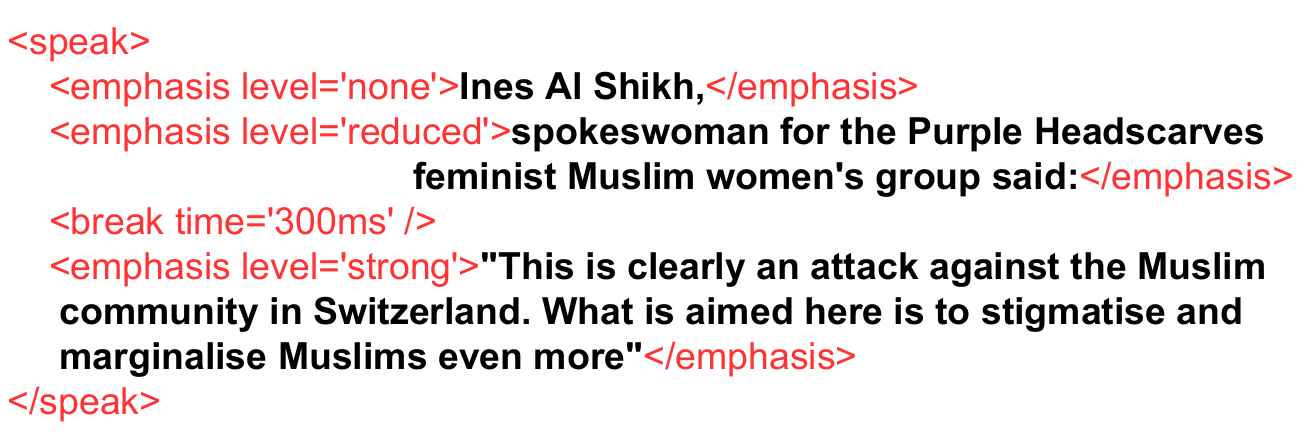}
    \caption{We use SSML to explicitly define emphasis and silences in quote paragraphs, which is taken into account by the WaveNet speech engine.}
    \label{fig:quote_ssml_example}
\end{figure}

Once the podcast's segments are drafted, we proceed to generating the audio using Google's Text-To-Speech API\footnote{https://cloud.google.com/text-to-speech/}, which gives access to several WaveNet-based voices \cite{oord2016wavenet}.

Of the 90+ voices available, we choose three based on recommendations from the API's documentation:
\begin{itemize}
    \item \textbf{Voice 1} is assigned to \texttt{en-US-Wavenet-J}, a male-identified voice responsible for voicing the introductory summary, and the paragraphs answering questions in the Q\&A Session. Voice 1 is the default voice, also responsible for greetings and transitions (e.g. ``Next up, ...'')
    \item \textbf{Voice 2} is assigned to \texttt{en-US-Wavenet-H}, a female-identified voice responsible for voicing questions in the Q\&A Session.
    \item \textbf{Voice 3} is assigned to \texttt{en-US-Wavenet-D}, a male-identified voice responsible for voicing quotes in the final portion of a segment.
\end{itemize}

In the example podcast segment of Figure~\ref{fig:example_segment}, the text is colored according to the assigned voice, with blue, red and green representing voices 1, 2 and 3 respectively. We note that we select WaveNet voices with an American English accent, as our usability studies were performed with crowd-workers mainly from the United States. However, WaveNet supports English accents from other regions, such as Australia, Great Britain and India, which could be used to further customize the experience.

The WaveNet model automatically applies some intonation and emphasis based on patterns learned in its training data (e.g., output a short silence after a comma). On top of that, the WaveNet API supports the Speech Synthesis Markup Language (SSML) \cite{taylor1997ssml}, allowing the explicit placement of emphasis, prosodic and silence markers. We use regex-based rules to apply SSML markers on a subset of paragraphs, as we found that the markers help reduce monotony. In Figure~\ref{fig:quote_ssml_example}, we illustrate how SSML is applied to paragraphs that match the quote format, in which we vary the emphasis between the author's name, their introduction, and the quote itself.

\subsection{Live Q\&A Response}
\label{section:live_qa}

Listeners have the option to interact with the podcast by asking free-form questions during a segment, which can be enabled by clicking on a microphone button.
Modern web browsers have an integrated speech recognition engine using a standardized interface\footnote{https://developer.mozilla.org/en-US/docs/Web/API/SpeechRecognition}, which we use to transcribe the user's speech into text.

We use the same QA model used in Section~\ref{section:gen_qa_session} to build the Q\&A graph to answer user-prompted questions.
Specifically, we use the QA model to see if any of the segment's paragraphs contain an answer to the question.
Importantly, the QA model is \textit{extractive}, and can only answer questions that have answers in one of the source news articles. The QA model cannot combine information and synthesize text, which could be required in some open-ended questions.

Running the neural network-based QA model introduces a latency of 4 seconds on average. Therefore, when the user submits a question, the system notifies the listener with the following message: ``I'll look into that, give me a moment.''

If the QA model returns an answer, it is used to fill in a reply template. For instance, one of the participants in Usabilibity Study B asked the question: ``Where are Rohingya refugees from?'' and received the following answer:
\begin{quote}
    ``I think the answer is \textbf{Myanmar}, I got it from the following paragraph. \textbf{About 170 Rohingya refugees were told they will be forcibly deported back to Myanmar where they had previously fled genocidal human rights abuses.}''
\end{quote}

On the other hand, if the QA model is not confident that any paragraph contains an answer, the system informs the listener with:
\begin{quote}
    ``Sorry. I couldn't find the answer. If you rephrase I will try again. Otherwise I'll keep walking you through the story.''
\end{quote}

\noindent
The podcast then resumes from the interruption point, skipping any sentence that it was in the midst of.

In order to understand whether a listener would be interested in interrupting the podcast with custom questions, we design a study (Section~\ref{sec:study_interaction}) centered around this interaction, evaluating how often interruptions occur, whether participants are satisfied with the returned answers, and analyzing the types of questions participants submit.

\subsection{Implementation Details}

We developed the system back-end in \texttt{Python}, and the interface front-end in \texttt{HTML}, \texttt{CSS}, and \texttt{JavaScript}.
The front-end communicates with the back-end using an API built with the Flask library\footnote{https://github.com/pallets/flask}.

For the summarizer, the question generator, and the question answering models, we used HuggingFace's Transformer library\cite{wolf-etal-2020-transformers}, more specifically leveraging the implementation of models for the PyTorch library. For the summarizer, we used a pre-trained model \cite{Zhang2020PEGASUSPW}, but we trained the Question Generator and Question-Answering models ourselves, using GPT-2 \cite{radford2019language} and Roberta-Large \cite{liu2019roberta} architectures respectively. We use a Tesla V-100 Nvidia GPU to run the models and to generate summaries and questions. We estimate that training the models required between 2 and 4 GPU days.

As news articles are added to our overall collection, we continuously run the pipeline described in \S\ref{section:identify_sections}-\ref{section:gen_qa_session}. A typical segment is based on 6 source news articles, with 50-100 paragraphs of content. Generating the Q\&A Session is the most computationally expensive component of the pipeline, taking on average around 50 seconds. The two compute-heavy steps are generating the questions (on average 9 seconds), and running the QA model to build the Q\&A graph (on average 40 seconds), while all other steps are almost instantaneous.

In order to have granular alignment between the transcript and the audio, we generate an individual audio file for each sentence of the podcast, allowing us to add sentences gradually into the transcript panel of the graphical interface. We store audio files in the standard OGG file format \cite{Pfeiffer_theogg}, an open-source audio file format compatible with \texttt{HTML5} and most modern web browsers.

\section{Usability Study A: Narration Style}
\label{sec:study_structure}

\subsection{Study A Design}

We perform a usability study focused on evaluating whether the automatic organization and generation of content described in \S\ref{section:identify_sections}-\ref{section:gen_qa_session} produces coherent and interesting news podcast segments.

Keeping other components fixed (such as the interface, the stories described, the speech engine), we compare four settings: two fully written by hand (Baseline and Reference), and two composed by an automated algorithm (QA Best and QA Rand). In order to ensure a fair comparison, all settings must generate segments of the same target length. For the study, we set the target at 200 words, for an expected segment duration of 90 seconds, assuming a speech rate of 120-150 words per minute.

\textbf{Baseline}. We randomly select one of the segment's source articles, and select the first $k$ sentences such that there is a total of at least 200 words. We do not generate questions, and the entire segment is voiced by Voice 1. This setting is similar to existing automated podcasts, because the chosen content is read without being adapted for the audio format.

\textbf{QA Best}. We generate the segment following the procedure described in Section~\ref{sec:system_description}. Specifically, questions are automatically generated, and the selection and ordering of questions follows the algorithm illustrated in Figure~\ref{fig:qa_session_pseudocode}.

\textbf{QA Rand}. Similar to QA Best, questions are automatically generated; however, the selection of questions as well as the answering paragraph are randomized (line 6 and 8 in Figure~\ref{fig:qa_session_pseudocode} are replaced by uniform random sampling).

\textbf{Reference}. One of the authors of the paper, with a background in journalism, wrote podcast segments in the Q\&A format. The process involved reading one or more of the source articles, and a combination of copying content and writing novel sentences (e.g., the questions).

Comparing these four settings, we aim to answer the following three research questions:
\begin{itemize}
    \item \textbf{RQ1}: Is the Q\&A format adequate for news domain podcasting, compared to top-down article reading? (Baseline vs. Reference)
	\item \textbf{RQ2}: How do automated Q\&A Sessions compare to hand-written Q\&A Sessions? (QA Best vs. Reference)
	\item \textbf{RQ3}: How effective is the ordering algorithm in Figure~\ref{fig:qa_session_pseudocode} at creating a relevant and coherent Q\&A Session? (QA Best vs. QA Rand)
\end{itemize}

Appendix~\ref{appendix:example_podcast_segments} gives a side-by-side comparison of the settings for one of the segments included in Usability Study A.

\subsection{Study A Specifics}

\begin{table*}[ht!]
	\caption{Detail of the seven podcast segments of Usability Study A. Participants selected 5 preferred segments (out of 7 possible), and were randomly assigned to a setting: Baseline, QA Rand, QA Best, or Reference. \textit{\#Source} is the number of source news articles in the segment, \textit{\#Part.} the number of participants that selected each segment, \textit{$\star$I} the average \textit{How Interesting} rating, \textit{$\star$C} the average \textit{How Coherent} rating, and \textit{\#QA} the number of question/answers spoken in the segment (Baseline did not have Q\&A). Methods are each assigned a symbol used to indicate a statistically significant difference ($p<0.05$) with other methods.}
	\label{table:study_a_detail}
\includegraphics[]{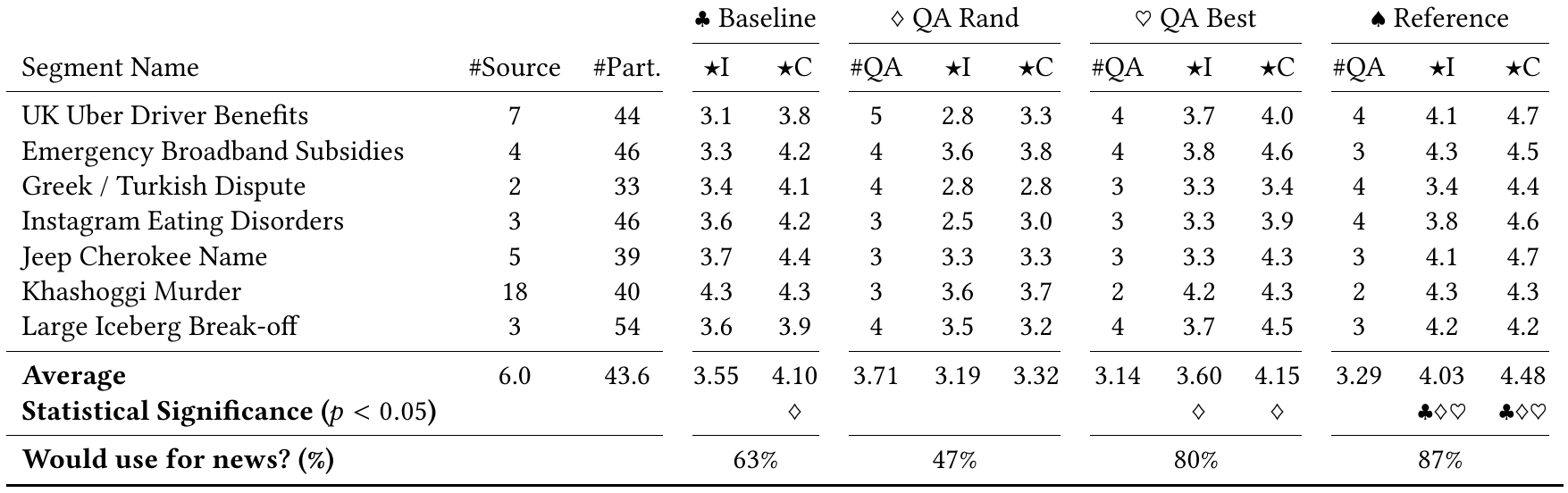}
\end{table*}

We recruited participants through Amazon Mechanical Turk, a crowd-sourcing platform. Crowd-workers participated from their computer, which represents a different setting than an audio-only setting such as listening to a podcast while driving or cooking. This allows us to first evaluate the system with participants whose main focus is on the podcast content; we leave evaluating the podcast in other settings to future work.

We designed the study to run over fifteen minutes, with the following procedure:
\begin{enumerate}
    \item \textbf{Introduction}: Participants were introduced to the task, gave their consent to participate, and viewed slides outlining the capabilities of the podcast interface (i.e. how to pause, open the transcript, etc.).
    \item \textbf{Sound Check}: Upon clicking on a button, an audio message would be generated: ``Click on button C''. The final letter was randomized, and the user had to click on the correct letter's corresponding button (out of four). This was intended to check that participants had sound enabled.
    \item \textbf{Segment Selection}: Participants selected 5 segments out of a total of 7 available, detailed in Table~\ref{table:study_a_detail}. This choice was intended to allow participants to customize podcast content, and reduce the chances they had to listen to topics they were not interested in.
    \item \textbf{Podcast Listening}: The participant was randomly assigned to one of the four settings for the entire podcast. Without pausing, the podcast had an expected duration of 7 and 8 minutes. After each of the five segments, participants had to complete a short segment satisfaction form, detailed below.
    \item \textbf{Post-completion questionnaire}: participants completed the study with a satisfaction questionnaire.
\end{enumerate}

Importantly, in this study's setting, we hid the microphone button in the interface, preventing participants from interrupting with their own questions. This was intended to focus this study on the podcast content itself. Study B is described in Section~\ref{sec:study_interaction} focuses on listener interactions.

We hand-selected the topics of the seven segments that participants chose from, creating a diverse set ranging from science (e.g., Large Iceberg Break-Off) to international affairs (e.g., Greek / Turkish Dispute). For the study, we specifically avoided segments that were predominantly political.

There were three main sources of signal we recorded:
\begin{enumerate}
    \item \textbf{Interaction Log}: recorded how often the participant paused the podcast, skipped a segment or opened the transcript.
    \item \textbf{Segment Satisfaction Form}: upon completing each segment, participants answered three questions: \textit{Already Know}: Did you know about this segment already? (yes/no), \textit{How interesting?}: How interesting was the segment? (1 to 5 stars), \textit{How Coherent}: How coherent was the segment? (1 to 5 stars).
    \item \textbf{Post-Completion Questionnaire}: upon completing the podcast, participants answered two questions: (1) a binary question: Would you use this system to get updated on the news? (yes/no), and (2) an optional free text feedback on the overall experience.
\end{enumerate}

We used Amazon Mechanical Turk to recruit participants, restricting the task to workers in English-speaking countries having previously completed 1600 tasks (HITs) and an acceptance rate of at least 97\%. Each participant was paid a flat rate of \$2.50 with the study lasting a maximum of 15 minutes. Appendix~\ref{appendix:usability_study_instructions} provides instructions given to participants.

\subsection{Study A Results}

We ran the study with a total of 61 participants; 16 were assigned to the Baseline, and 15 each to QA Best, QA Random and Reference.

Even though segment names were listed in shuffled order across participants during segment selection, there were some preferences for some segments, with the most popular being about a Large Iceberg Break-Off and the least popular the Greek / Turkish Dispute.

With regard to the interaction log, 56\% of participants opened the transcript at least once, 22\% of participants paused the podcast at least once, and 5\% skipped one or more segments, with no significant difference across settings.

The Segment Satisfaction Form results are compiled in the Table~\ref{table:study_a_detail}.
Because we are comparing all pairs of four total settings, we performed a Kruskal-Wallis test \cite{kruskal1952use} with a Dunn post-hoc test \cite{dunn1964multiple} for statistical significance between pairs of conditions.

The hand-written Reference condition outperforms all other methods significantly, both in how interesting and how coherent the podcast segments are. QA Best achieved the second-best scores in both metrics, narrowly above the Baseline, without statistical difference between the two settings. Finally, QA Random comes in last for both metrics, with a large average drop of 0.8 stars on coherence compared to QA Best, shining a light on the success of QA Best's selection and ordering of questions.

We measure Pearson cross-correlation between \textit{Already Know} and \textit{How Interesting}, and find a positive correlation between the two of 0.24: if a listener is already aware of a particular topic, they are likely to rate it as more interesting. The correlation between \textit{Already Know} and \textit{How Coherent} is small (around 0.1), but the correlation between \textit{How Interesting} and \textit{How Coherent} is highest at 0.5: indeed, the podcast's coherence can be seen as a prerequisite for it to be interesting.

Fifty-six of the sixty-one participants opted to give general feedback via an open-ended text box. We tagged the responses into major themes:
\begin{itemize}
    \item 30 participants expressed positive interaction with the system (e.g., \textit{Nice work, I actually enjoyed listening to this. The female voice with questions was a nice touch. Good job.}, Participant A4 - Reference)
    \item 11 participants mentioned that the speech quality was problematic, (e.g., \textit{I would like a more natural voice if possible.}, Participant A42 - Baseline), with 4 participants mentioning a voice in particular (e.g. \textit{I liked it [...] but didnt like the guy's voice}, A16 - Baseline)
    \item 9 participants mentioned the interface was easy to use (e.g., \textit{I think the system here is a very simple but easy to use interface that even less tech-savvy users can easily use to obtain the latest news.}, A13 - QA Best)
    \item 3 participants mentioned the generated speech was too fast (e.g., \textit{Some of the speech was too fast to understand.}, Participant A28 - Reference), 3 participants wished they could select from more segments (e.g., \textit{their was not to many articles to choose from.}, A38 - Reference), 3 participants mentioned the transcript was useful (e.g., \textit{The live transcript feature worked well, as you'd expect for computer-generated speech)}, A39 - QA Rand).
\end{itemize}

With regards to the research questions we posed:
\begin{itemize}
    \item \textbf{RQ1}: Yes, the Q\&A format is adequate for news domain podcasting, with participants significantly preferring it over a baseline of reading unaltered news articles.
    \item \textbf{RQ2}: Automated Q\&A sessions do not perform as well in terms of coherence and interestingness when compared to hand-written references, but 80\% of QA Best listeners said they would use the system to listen to the news in the future.
    \item \textbf{RQ3}: The ordering choices based on the greedy traversal of the QA Graph make a significant difference when choosing questions compared to random sampling in terms of coherence and interestingness.
\end{itemize}

\section{Usability Study B: Interaction}
\label{sec:study_interaction}

\subsection{Study B Design}
\begin{figure}
    \centering
    \includegraphics[width=\linewidth]{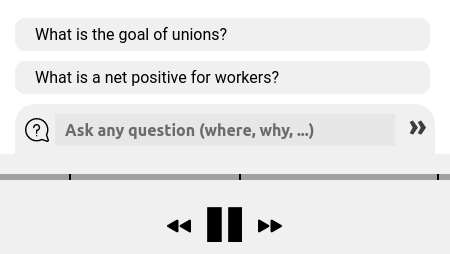}
    \caption{For the Interaction Study, we modify the interface, removing the microphone option (shown in Figure~\ref{fig:newspod_interface_overview}), allowing the participant to either type a question, or click on two recommended questions.}
    \label{fig:interface_questions}
\end{figure}

In the Interaction Study, we set the content for all participants to the QA Best setting, and focus on measuring what kinds of interruptions, if any, listeners of news podcasts might initiate.

Even though there is evidence that using speech recognition can ease user participation \cite{Karl1993SpeechVM}, there are limitations to speech input interfaces \cite{Shneiderman2000TheLO}, particularly in the case of crowd-workers who may be in environments in which they are not willing to speak out loud. Preliminary experiments with making NewsPod's microphone button visible confirmed that a vast majority of crowd-workers might not want to enable their microphone for a 15-minute study. We therefore chose to replace the microphone-enabled interruptions with a text-box for the purpose of the study, allowing participants to type their questions instead of speaking them.

We also added two recommended questions to each podcast segment. Participants could choose at any point during a podcast segment to type their own question, or click on a recommended question. The graphical change to the interface is illustrated in Figure~\ref{fig:interface_questions}.

All participants are assigned to the QA Best segment content; however, they are randomly split across two settings:
\begin{itemize}
    \item \textbf{With Break}: At the end of each segment, the following break paragraph is added: \textit{We're wrapping up this story, if you have a question, now is a good time to ask. Otherwise, we'll be moving on to the next story.} Each of the two sentences is followed by a five second break. This adds a 15-second period dedicated to listeners being able to ask a question at the end of each segment.
    \item \textbf{Without Break}: The podcast segments do not include any built-in breaks for participants to ask questions.
\end{itemize}

In both cases, participants can ask a question or click on a recommended question at any point in the podcast, but in the \textbf{With Break} setting, an additional period is added specifically to encourage participants to ask questions.

We aim to answer the following research questions:
\begin{itemize}
    \item \textbf{RQ4}: Are news podcast listeners interested in interacting with the podcast by asking questions?
    \item \textbf{RQ5}: Does including breaks specifically designed for listeners to ask questions help increase user interaction?
    \item \textbf{RQ6}: Are current automated question-answering models equipped to answer news listeners' questions?
\end{itemize}

\subsection{Study B Specifics}

This study was modeled on the previous one, with participants going through an introduction, a sound check, the segment selection, podcast listening, and a post-completion questionnaire.

Because we ran this study two weeks after Study A, we generated a new set of 6 QA Best podcast segments in order to maintain high content relevance for the participants. Details of the six segments options are given in Table~\ref{table:interaction_questions}.

In order to keep the study in the fifteen minute window while accommodating for interruptions from the listener, listeners were asked to choose four segments (out of six total) instead of five (out of seven total) as in Study A.

We recruited participants on Amazon Mechanical Turk with the same remuneration and the same selective filters on eligible crowd-workers as in Study A. Crowd-workers that had already completed Study A were ineligible for Study B. In total, 40 participants completed User Study B, with 20 participants in the Without Break setting, and 20 in the With Break setting.

We modified the post-completion questionnaire to focus on the interaction from the participant: for each question asked (recommended and typed) that the system attempted to answer, the participant was asked \textit{How was the system's answer?}, and given three options: \textit{I don't remember}, \textit{Irrelevant/Confusing}, and \textit{Good/Relevant}.

\subsection{Study B Results}

\begin{table}[]
    \caption{Results of Usability Study B on the six segments, with 40 participants randomly assigned to Without Break (20) or With Break (20) settings, reporting: \textit{\#Part.} the number of participants that selected each segment, \textit{Rec}, \textit{Own} the average number of recommended questions clicked and own questions typed, and \textit{Any} the percentage of participants that asked at least one question. In total row, * indicates a statistically significant difference between settings (p<0.05).}
    \label{table:interaction_questions}
    \resizebox{0.47\textwidth}{!}{%
	\begin{tabular}{lccccccc}
	 & & \multicolumn{3}{c}{\textbf{Without Break}} & \multicolumn{3}{c}{\textbf{With Break}} \\
    \cmidrule(r){3-5} \cmidrule(r){6-8}
	\textbf{Segment Name} & \textbf{\#Part.} & \textbf{Own} & \textbf{Rec} & \textbf{Any} & \textbf{Own} & \textbf{Rec} & \textbf{Any} \\
	\cmidrule(r){1-2} \cmidrule(r){3-5} \cmidrule(r){6-8}
	Amazon Union & 37 & 0.4 & 0.5 & 47\% & 0.3 & 0.8 & 60\% \\
	Bethesda Acquisition & 33 & 0.2 & 0.4 & 41\% & 0.2 & 0.9 & 69\% \\
	Rohingya Crisis & 20 & 0.4 & 0.2 & 33\% & 0.6 & 0.6 & 64\% \\
	Senegal Jailed Leader & 15 & 0.0 & 0.5 & 33\% & 0.2 & 1.0 & 67\% \\
	Swiss Burqa Ban & 23 & 0.2 & 0.4 & 46\% & 0.1 & 1.0 & 80\% \\
	Tesla Self Driving Ban & 32 & 0.1 & 0.2 & 17\% & 0.2 & 0.7 & 57\% \\
	\midrule
	\textbf{Total} & 160 & 0.9 & 1.5 & 55\% & 1.1 & \textbf{3.2*} & \textbf{85\%*} \\
	\bottomrule
	\end{tabular}
	}
\end{table}

Participants completed a Segment Satisfaction Form identical to the one described for Study A. Podcast segments in Study B were judged to be roughly as interesting and coherent as in Study A, averaging 3.9 and 4.1 stars respectively (vs. 3.6 and 4.2 in the Study A).

Table~\ref{table:interaction_questions} summarizes statistics on questions asked by participants, both clicked recommendations and typed questions. We find that in the condition without breaks, roughly half the participants asked a question, while in the condition with breaks, the proportion increases to 85\%.

Including breaks also significantly increases the average number of recommended questions clicked (from 1.5 to 3.2) and in a non-significant way the number of questions typed (0.9 to 1.1).

\textbf{RQ4, RQ5}: When constructing the podcast with breaks, a large majority (85\%) of podcast listeners asked at least one question throughout the podcast; substantially more than when no break was offered.

Next, we report on participant satisfaction of the quality of answers for selected questions. For those questions that were generated by the system, participants were generally satisfied with answers, with 79\% of the answers rated \textit{Good/Relevant}.

\textbf{RQ6}: On the other hand, participants were mainly dissatisfied with the automatically generated answers to their own typed questions: 76\% of the answers were rated \textit{Irrelevant/Confusing}. This result reveals the limitation of applying an extractive question answering system to open question answering.

To understand further what types of questions participants formulated, and why the live Q\&A system might be inadequate, we analyzed and categorized the 57 typed questions:
\begin{itemize}
    \item \textbf{Factoid Questions} make up 36\% of questions, asking for a specific detail that can be extracted from one of the source articles. (e.g., \textit{what percent of amazon workers belong to a union?}, Participant B18 - Without Break)
    \item \textbf{Synthesis Questions} make up 18\% of questions and require compiling and summarizing several elements to answer (e.g., \textit{how have people reacted to the ban?}, B6 - Without Break)
    \item \textbf{Encyclopedic Questions} make up 14\% of questions and ask for information most likely present in an encyclopedia such as Wikipedia, or in a knowledge base (e.g., \textit{What type of government does Senegal have?}, B23 - With Break)
    \item \textbf{Clarification Questions} make up 13\% of the questions, asking to define a term in the podcast (e.g. \textit{Who is Bethesda, again?} B24 - Without Break)
    \item \textbf{Prediction Questions} make up 9\% of questions, asking about hypothetical future events (e.g., \textit{When will Tesla release a new edition?} B12 - With Break)
    \item \textbf{Rhetorical Questions} make up 5\% of questions, making a statement not requiring an answer (e.g., \textit{What about driving safety?} B14 - With Break)
    \item \textbf{Self Relevance questions} make up 5\% of questions, relating the event to the listener (e.g. \textit{How can I join in this union?}, B36 - With Break)
\end{itemize}

Even though news content is heterogeneous and news articles potentially contain answers to a wide variety of question types, the QA system we operate is mainly equipped to deal with \textbf{Factoid Questions}, since they represent the vast majority of questions present in QA datasets used to train the model. This mismatch between participant expectations and system capabilities leads to the dissatisfaction with most answers to typed questions.

Further analysis reveals that the inadequacy of the QA responses lead to a reduction in participant interaction with the system. Even though the model is able to satisfactorily answer comprehension questions 79\% of the time, participants judged that the last question they asked had a satisfying answer only 57\% of the time. We interpret this as evidence that inadequate responses from the QA system can discourage participants from asking further questions.

\section{Discussion}

Overall, news podcasts in the conversational format were largely enjoyed, with almost 90\% of participants noting they would use the hand-written Reference system again, compared to 63\% for the Baseline podcast based on voicing unmodified news articles.
Currently, automating the podcast by using neural network-based methods to generate summaries, questions and ordering the content causes a loss in quality when compared to manual curation; however, most listeners still enjoy the podcast, with 80\% of \textit{QA Best} participants stating they would use the podcast again. Continual improvements in NLP technology will gradually reduce this gap, facilitating the construction of automated audio content for the news and other domains.

Listeners of podcasts are interested in interacting with audio content, with a majority of listeners interrupting the podcast to ask a question when the option is available. Including appropriate silences and invitations to participate in the podcast script is essential to increasing interaction, as in our study this increased the proportion of participants asking questions from 55\% to 85\%. Though powerful, current question answering systems are still too limited and specific to deal with the diversity of questions posed by listeners, and further engineering is required to enable satisfying interactions with the listener.

The usability studies of our system illustrate the potential of audio interfaces to increase in-depth news engagement: in a few minutes and potentially while performing another activity, a listener can obtain detailed information about topics of their choice, from a potentially diverse set of sources. At any point, the listener can join the conversation and ask questions, clarifying unclear aspects, and steering the discussion dynamically.

\section{Limitations}

\textbf{At the desk setting.} In our current usability studies, our participants were predominantly using laptop computers, most likely at a desk. A desk environment is advantageous, as it provides a better sound system, but is not representative of the large portion of environments for podcast listening \cite{newman_2020}. Further studies of the usability of NewsPod in the car, or while walking, cooking, or doing other everyday activities could shine light on other advantages and limitations of the system.

\textbf{Unevaluated Podcast Elements.} Some of the design decisions in the podcast, such as the choice to include quotation paragraphs, or the wording in the greeting and closing sentences, were not explicitly evaluated in the usability studies. Therefore interpretation of the system's success and limitations should be limited to the main elements evaluated, such as the format of the Q\&A session.

\textbf{Amazon Mechanical Turk population bias.} Even though recruiting through a crowd-sourcing platform typically leads to more representative participants than on-site recruitment (typically, undergraduate students at a university) \cite{Berinsky2012EvaluatingOL}, the crowd-sourced population differs from the U.S. population on many metrics (e.g., age, gender, income level) \cite{Ipeirotis2010DemographicsOM} which would likely have an impact on our results.

\textbf{Imperfect Speech and Q\&A systems.} Recent work \cite{Cambre2020ChoiceOV} has shown that modern text-to-speech has approached, but not achieved, recorded human speech in quality, and around 15\% of our participants stated in their open feedback that low speech quality was noticeable. Similarly, our Q\&A system was not able to provide satisfying answers to a majority of the participants' questions. These imperfect components most certainly had a negative impact on participants' evaluation of the system; however, we did not measure the extent of these effect.

\textbf{Factuality of Generated Text.} Several components in the creation of a NewsPod podcast, such as the Summarization and Question Generation modules, generate novel text. This novel text is included in the podcast script without human verification, and could include factually incorrect information, potentially misleading or confusing the listener. Recent progress at detecting factual inconsistencies indicate a path towards consistent text generation \cite{laban2021summac}, but for now the use of generated text in applied settings such as NewsPod is limited.  

\textbf{Lack of Editorial Judgement.}  High-quality journalism includes the exercise of judgement, an attempt at balance, and careful attention to sources. The algorithms presented here do not attempt to exercise such judgement; as just one example, only one quotation is presented, without the standard journalistic practice of an answering quotation. Future work should incorporate these considerations.

\section{Future Work}
\label{sec:future_work}

\textbf{Adapting to User Interactions.} When a listener asks a specific question, the system attempts to answer, and then immediately resumes the previous planned segment. However, it is likely that an interaction should lead to changes later on in the podcast, as it might have answered questions further down in the segment, or should lead to follow-ups that would interest the listener more.

\textbf{Customizing to Repeat Listeners.} A news podcast is an opportunity for a conversational interaction with the listener. The ability to remember prior podcasts, and integrating them seamlessly into the generation of future podcasts could lead to exciting interactions: ``\textit{Hey Robin, we talked about Brexit last Wednesday, and here's the update since then...}''.

\textbf{Including Sound and Music}. Studies have shown the effectiveness of musical communication in cinema \cite{lipscomb2005role}, and more recent work \cite{Steinhaeusser2021ComparingAR} has shown that adding music and sounds to audio-books increases transportation and fear effects in listeners, both for human-recorded and automated audio-books. For automated news podcasts, music and sounds could be effective communication tools, and help soften perceptibility of automated speech flaws.

\section{Conclusion}

This paper introduced NewsPod, an interactive and automated news podcast. NewsPod organizes news content into a simulated conversation in which several automated voices ask and answer questions about a topic. The listener can at any point join the conversation by asking a question, which the system attempts to answer automatically.
We evaluated the Q\&A narrative style in NewsPod through a usability study and found it led to participants finding the content more enjoyable and less monotonous, with 80\% of the participants saying they would use the system again to stay informed on the news. A second usability study, centered on participant interactions, reveals that when appropriate breaks are included in the podcast, 85\% of listeners ask questions and interact with the podcast. Participant-prompted questions are found to be diverse, with a majority tripping up current neural network-based question-answering models, reflecting current limitations in natural language processing. This alternative approach to news consumption may eventually serve to broaden the base of users who engage deeply with the news.

\begin{acks}
We would like to thank Katie Stasaski, Forrest Huang and Eldon Schoop and the IUI reviewers for their valuable feedback, as well as Ruchir Baronia for development of early prototypes. This work was supported by a Microsoft BAIR Commons grant
as well as a Microsoft Azure Sponsorship.
\end{acks}

\bibliographystyle{ACM-Reference-Format}
\bibliography{bibliography}

\appendix

\section{Usability Study Instructions}
\label{appendix:usability_study_instructions}

The following instructions were given to all participants. Participants were expected to accept the terms before moving on to the study:
\begin{itemize}
    \item The entire HIT should take no more than 15 minutes:
    \item You must use Google Chrome or Mozilla Firefox for this task.
    \item You will go through a tutorial about the interface.
    \item Listen to a news podcast for 7 minutes. You must have sound enabled to hear the podcast.
    \item You can leave at any point but will not complete the HIT.
    \item You can complete this task at most once.
    \item If you have a question/problem, contact us at X@email.com
    \item By clicking ``I Accept'', you agree to our terms. Namely: we will use the data will collect for research purposes. We do not collect any personally identifiable information. You agree to performing the task to the best of your ability.
\end{itemize}

\section{Example Podcast Segments}
\label{appendix:example_podcast_segments}
\begin{figure*}
    \centering
    \begin{subfigure}[b]{0.475\textwidth}
        \centering
        \includegraphics[width=\textwidth]{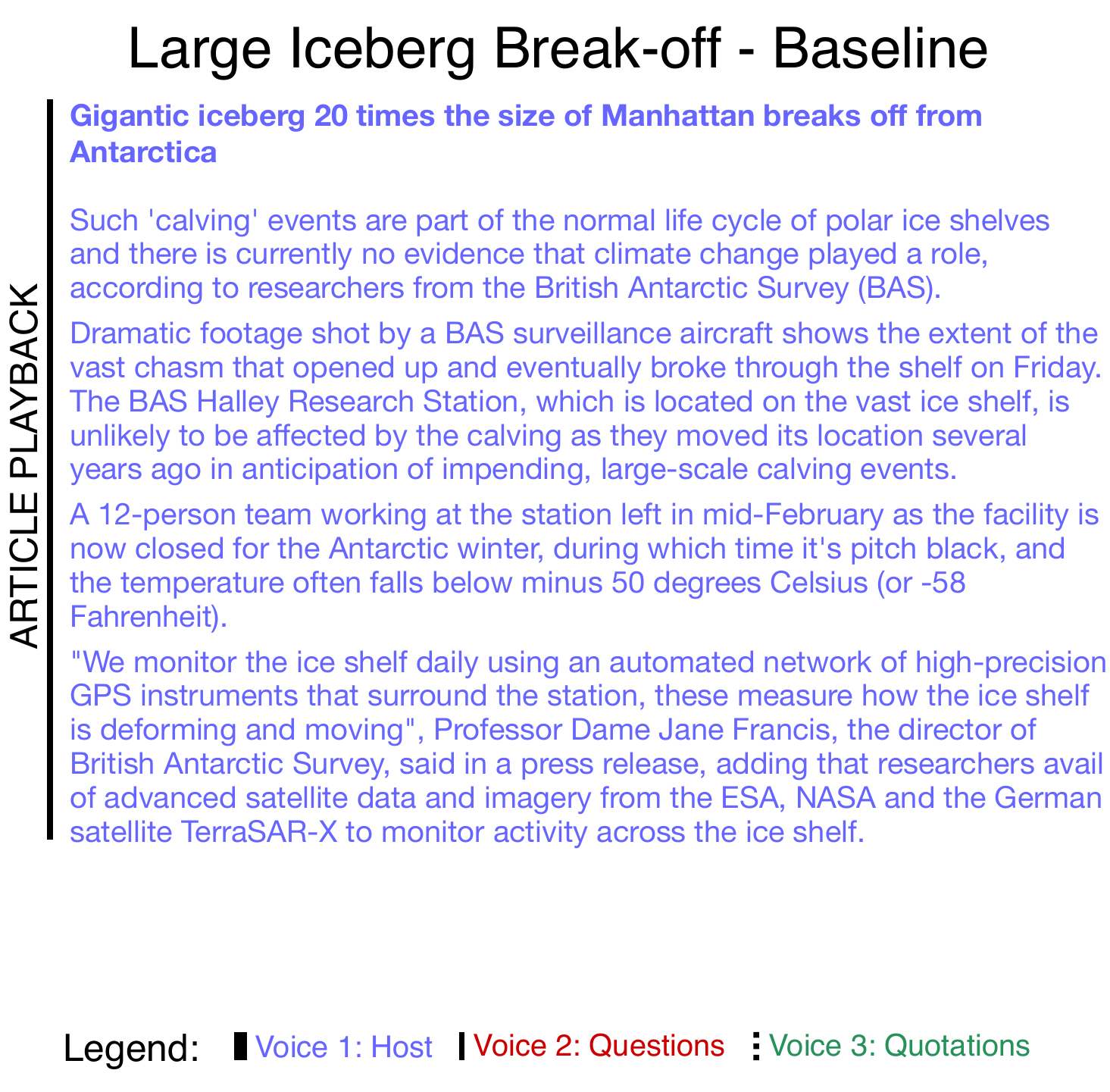}
        \caption[]{Baseline Setting}
        \label{fig:example_baseline_setting}
    \end{subfigure}
    \hfill
    \begin{subfigure}[b]{0.475\textwidth}  
        \centering 
        \includegraphics[width=\textwidth]{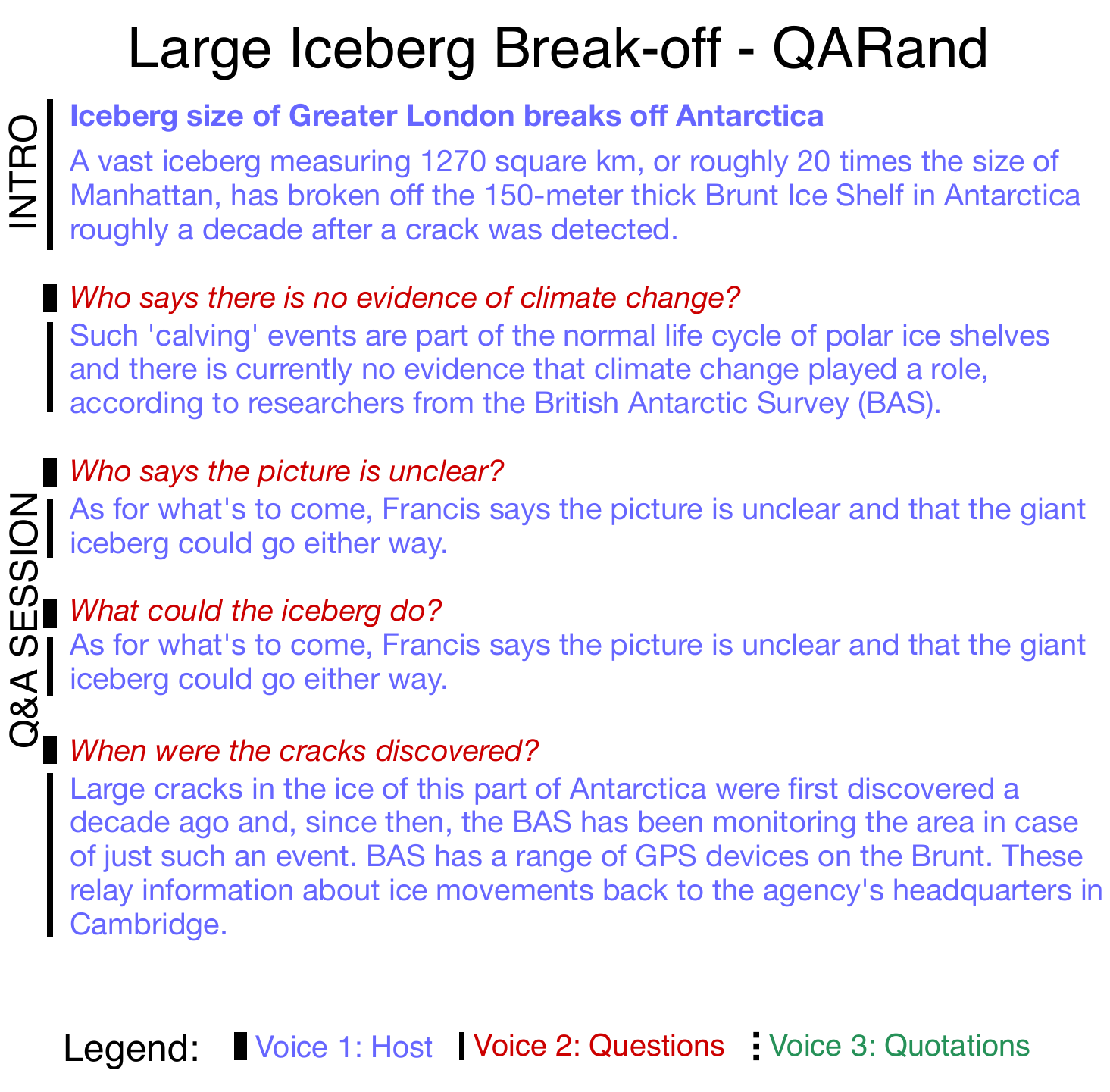}
        \caption[]{QA Rand Setting}
        \label{fig:example_qarand_setting}
    \end{subfigure}
    \vskip\baselineskip
    \begin{subfigure}[b]{0.475\textwidth}   
        \centering 
        \includegraphics[width=\textwidth]{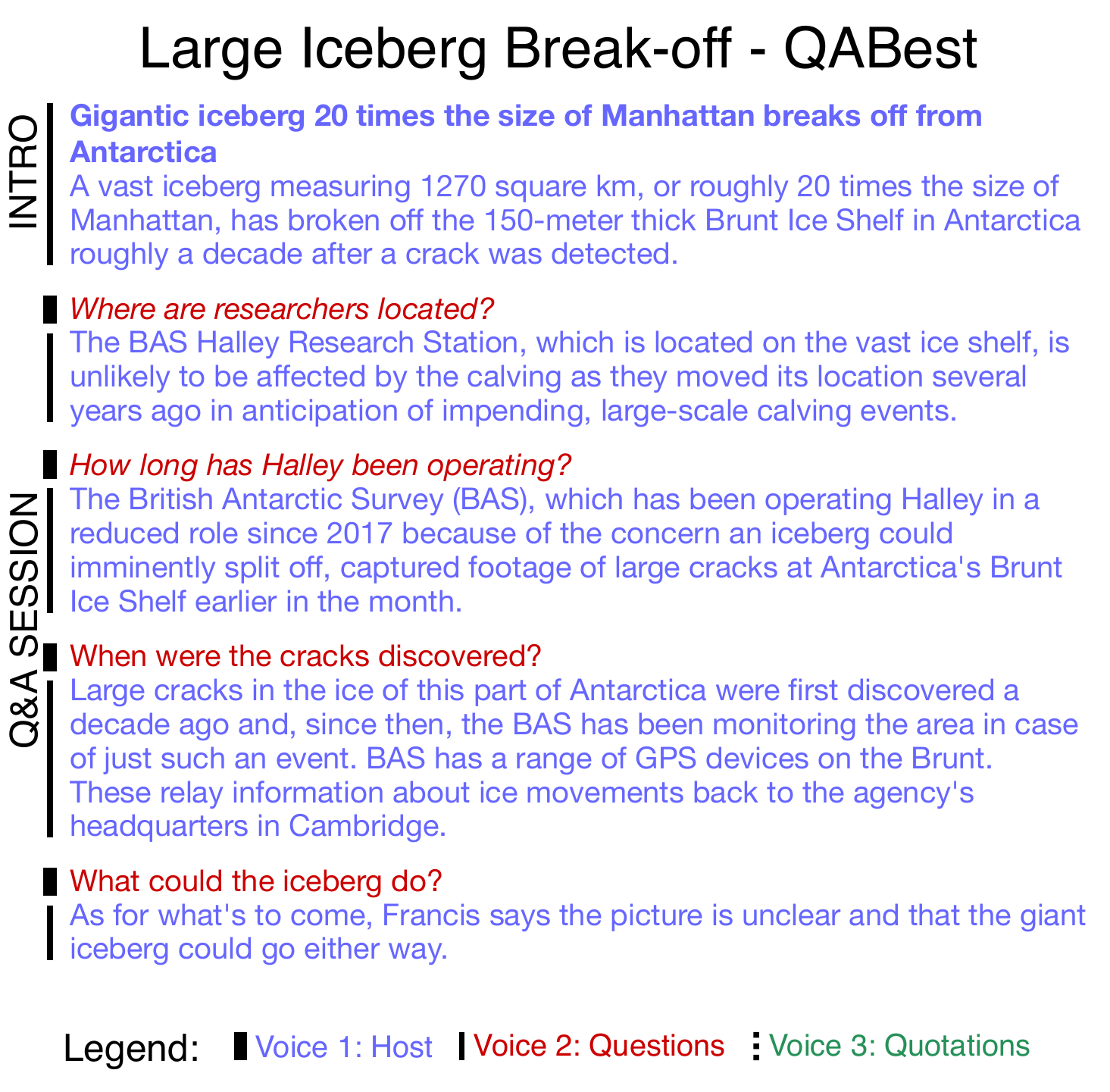}
        \caption[]{QA Best Setting}
        \label{fig:example_qabest_setting}
    \end{subfigure}
    \hfill
    \begin{subfigure}[b]{0.475\textwidth}   
        \centering 
        \includegraphics[width=\textwidth]{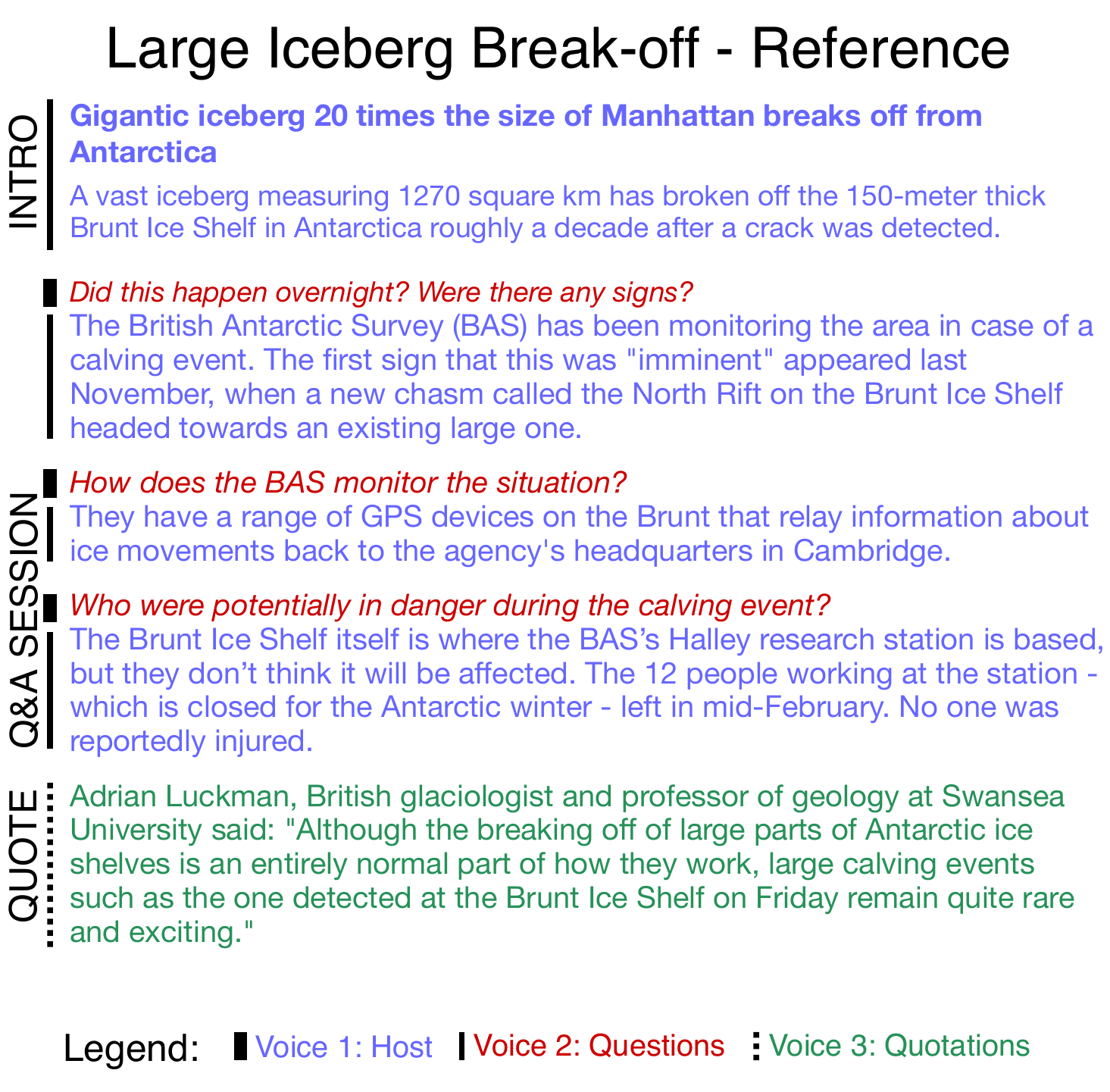}
        \caption[]{Reference Setting}
        \label{fig:example_reference_setting}
    \end{subfigure}
    \caption[]{Segments from the Large Iceberg Break-off topic from the four segments compared in Usability Study A: Baseline, QA Rand, QA Best and Reference.}
    \label{figure:four_segments}
\end{figure*}

In Figure~\ref{figure:four_segments}, we present the segments from the four settings we tested in Usability Study A (Baseline, QA Rand, QA Best, Reference) for the ``Large Iceberg Break-off'' segment. This can serve to understand differences across the settings that were tested. We note that for this particular topic, the QA Rand and QA Best automatic settings did not extract any quotations, and there no quotation paragraph was added to the segment.

\end{document}